\documentclass[
  aps,
  pre,
  reprint,
  superscriptaddress,
  floatfix
]{revtex4-2}

\usepackage{amsmath}

\DeclareMathAlphabet{\mathsf}{OT1}{phv}{b}{n}

\newcommand{\crossVorg}{\ensuremath{%
         \setbox0=\hbox{$V$}
        V \kern-\wd0{\raise.3ex\hbox{$\relbar$}}}}

\newcommand{\crossVxx}[2]{%
	{\setbox0=\hbox{$#1#2V$}
         \setbox1=\hbox{$#1#2$}
         \setbox2=\hbox{$#1V$}
         \dimen1=\wd0
	 \advance\dimen1-\wd1
         \raise.2\ht0\hbox{$#1#2$}\kern-.4\wd0}}

\usepackage{pifont}

\usepackage{multirow}
\usepackage{setspace}
\usepackage{enumitem}
\usepackage{verbatim}

\usepackage{amssymb}
\usepackage{amsbsy}
\usepackage[T1]{fontenc}
\usepackage{amsmath}
\usepackage{mathtools}

\usepackage{graphicx}
\usepackage{xcolor}
\usepackage{mathrsfs}
\usepackage[mathscr]{euscript}
\usepackage{tikz}
\usetikzlibrary{shapes}
\usepackage{cancel}
\usepackage[normalem]{ulem}

\usepackage[colorinlistoftodos]{todonotes}
\usepackage{verbatim}
\usepackage{latexsym}
\usepackage{textcomp}
\usepackage{bm}
\usepackage[Euler]{upgreek}
\usepackage{longtable}
\usepackage{subfigure}

\usepackage[colorlinks,allcolors=cyan!70!black]{hyperref}

\usepackage{epsfig}
\usepackage{dcolumn}

\usepackage{booktabs}

\newcommand{\qed}{\nobreak \ifvmode \relax \else
      \ifdim\lastskip<1.5em \hskip-\lastskip
      \hskip1.5em plus0em minus0.5em \fi \nobreak
      \vrule height0.75em width0.5em depth0.25em\fi}

\DeclareMathAlphabet\mathbfcal{OMS}{cmsy}{b}{n}
\DeclareMathAlphabet{\mathbfsf}{\encodingdefault}{\sfdefault}{bx}{sl}

\begin{document}

\setlength{\parindent}{1em}
\setlength{\parskip}{0pt}

\setlength{\abovedisplayskip}{6pt}
\setlength{\belowdisplayskip}{6pt}
\setlength{\abovedisplayshortskip}{4pt}
\setlength{\belowdisplayshortskip}{4pt}

% ---- your paper starts here ----

%\preprint{}

\title{Universality of the deswelling of tangentially active polymer chains in dilute solutions}


\author{Suryansh Tripathi}
\affiliation{Department of Chemical Engineering, Indian Institute of Technology (ISM) Dhanbad, Jharkhand 826004, India}
\author{Aritra Santra}
\email{aritrasantra@iitism.ac.in}
\affiliation{Department of Chemical Engineering, Indian Institute of Technology (ISM) Dhanbad, Jharkhand 826004, India}
 

\date{\today}

%\vspace{-20pt}

%\begin{figure}[ptbh]
% \begin{center}
% {\includegraphics*[width=!,height=7.5cm]{fig_TOC.pdf}}
% \end{center}
% \vskip-10pt
% \text{For Table of Contents use only}
%\end{figure}

%----------------------------------------------------------------------------------
\begin{abstract}
%----------------------------------------------------------------------------------
Dilute solutions of linear polymer chains with tangentially active monomeric beads are simulated using a Brownian dynamics (BD) algorithm over a range of solvent quality in the thermal crossover regime between $\theta$ and athermal solvents. The conformational changes with increasing P{\'e}clet number ($Pe$) (which is proportional to the strength of activity) suggest deswelling of the chains resulting in a collapse of the radius of gyration data to a random walk (RW) statistics at a unique value of $Pe$, independent of the solvent quality. The swelling behaviour of active polymers in the crossover regime relative to their size at the $\theta$ state is found to follow the same universal characteristics as that of passive polymer chains. Furthermore, based on polymer blob theory we present a novel scaling of the thermal blob size with tangential activity of the monomeric beads which leads to the definition of a renormalized solvent quality parameter for active polymers. Altogether, this work establishes a connection between the configurational properties of active polymers and scaling laws in polymer physics, which provides a useful framework to study the dynamics of activity induced motion of polymeric molecules for various applications in biophysics and other related areas.
%----------------------------------------------------------------------------------
\end{abstract}
%\keywords{}

\maketitle

\section{Introduction}
%%%%%%%%%%%%%%%%%%%%%%%%%%%%%%%%%%%%%%%%%%%%%%
Dynamics of active polymer solutions has recently emerged as an important focus of research primarily because of its relevance and applications in the field of biophysics, such as, in the study of replication of DNA~\cite{Alberts2015}, transcription of RNA~\cite{Mejia2015,Guthold1999}, or propulsion of actin filaments and microtubules~\cite{Liverpool2001,Mizuno2007,Bathe2008,Schaller2010,Nedelec1997}, all of which are governed by the dynamics of self-propelling bio-polymeric or filamentous molecules. Synthesis of artificial polymeric molecules with active monomers has also gained significant interests of the research community because of its utility in the development of micro-motors and advanced medicines for targeted drug delivery~\cite{Vutukuri2017,Hill2014,Nishiguchi_2018,Simmchen2016,Jain_AipAdv2022}. As compared to passive polymer solutions, which remain in thermal equilibrium unless acted upon by a large external force, active polymers are inherently out of equilibrium because of the active force components responsible for self-propulsion of the chain~\cite{Winkler_JCP2020,Demirel2010,Fang2019}. This leads to significant changes in the chain conformation and dynamics of active polymer solutions which are typically not observed for passive polymers~\cite{Bianco2018,Mousavi2021,Jain_Marco2022,Anderson2022,Panda_Macromol2025,Malgaretti2025,Panda_PRE2025,Philipps2022,Tejedor2024}. While extensive studies have been carried out to understand the static and dynamic properties of passive polymer solutions~\cite{Bird1987vol1,Bird1987vol2,Doi1988theory,Larson_PRE1997,Jain2015,Hsiao2016,Schroeder2018}, active polymer solutions are relatively less explored. Specifically, there are scaling relations with respect to the polymer chain size, monomer concentration, diffusivity and viscosity of passive polymer solutions which are used to characterize the viscoelastic properties of such systems and to understand the underlying physics~\cite{degennes1979,Smith1996,Sunthar2006,Jain_PRL2012,Pan2014}. However, such a framework is currently not well established for active polymer solutions. In this work, we have investigated these aspects by identifying a connection between the properties of tangentially active polymer chains and scaling laws derived from polymer physics in terms of a universal behaviour of the size of active polymer chains in the limit of dilute solutions.

It is well known that the radius of gyration or mean-squared end-to-end distance of polymer chains with asymptotically large molecular weight ($M_w$) follows a universal power-law scaling with respect to the molecular weight in the limit of $\theta$ and athermal solvent~\cite{degennes1979,Hayward1999}. Whereas, for polymer chains not having infinitely large molecular weight, the chain size shows a universal behaviour in the thermal crossover regime between $\theta$ and athermal solvent conditions with respect to a non-dimensional solvent quality parameter, $z$, defined as $z=k\,\hat{\tau}(T)\sqrt{M_w}$. Here, $\hat{\tau}(T)=(1-\Theta/T)$ is a temperature dependent term where $T$ is the solution temperature, $\Theta$ is the temperature at the corresponding $\theta$-solvent condition and $k$ is a constant dependent on the polymer-solvent chemistry~\cite{Miyaki1981,KumarPrkash2003,Sunthar2006,Santra2019,Santra2022}. Within the framework of polymer blob theory, the parameter $z$ is related to number of thermal blobs ($\mathcal{N}_T$) on a chain by $\mathcal{N}_T = z^2$~\cite{Dondos1996,degennes1979,Jain_PRL2012}. Specifically, the ratio of the radius of gyration of a polymer chain at a given solvent quality to its gyration radius at the $\theta$-solvent condition for a variety of polymer-solvent system is found to follow a universal curve as a function of $z$~\cite{KumarPrkash2003,Sunthar2006}. This curve is used to determine the effective solvent quality of the polymer solutions which ultimately controls the dynamics and viscoelasticity of the solutions~\cite{Sasmal2016,Santra2022}. Recent studies suggest that in the presence of excluded volume interaction, polymers with active force directed tangentially along the backbone of the chain show a coil-to-globulelike transition and enhanced diffusion with increasing P{\'e}clet number ($Pe$), where $Pe$ determines the strength of activity~\cite{Bianco2018,Tejedor2024,Fazelzadeh2023,Jain_Marco2022}. Such tangentially active polymer chains are useful model systems for studying translocation of filaments through pores, RNA transcriptions and DNA replications~\cite{Tan2023,Alberts2015,Mejia2015}. However, most of these studies focused on the dynamics of active polymers in dilute solutions under $\theta$ ($z=0$) or athermal ($z\rightarrow \infty$) limit, whereas, typically for solutions of polymer with finite molecular weight, the solvent quality remains in the crossover regime between $z=0$ and $z\rightarrow \infty$. The physics of tangentially active polymer chains in the thermal crossover regime remains largely unexplored. In this study, we implemented a Brownian dynamics (BD) simulation algorithm in the free-draining limit~\cite{KumarPrkash2003,Jain_PRL2012,Santra2022} to investigate the evolution of the size of tangentially active polymer chains in the thermal crossover regime and determined a scaling relation for the thermal blob size as a function of P{\'e}clet number to define a renormalized solvent quality parameter for active polymer solutions. In the subsequent sections we have presented details of the simulation methodology (in section~\ref{sec:SimMethod}), discussed the key results (in section~\ref{sec:results}) and finally, presented the summary and conclusion (in section~\ref{sec:summary}). 

%%%%%%%%%%%%%%%%%%%%%%%%%%%%%%%%%%%%%%%%%%%%%
\section{\label{sec:SimMethod}Simulation Methodology}
%%%%%%%%%%%%%%%%%%%%%%%%%%%%%%%%%%%%%%%%%%%%%%
Systems of dilute polymer solutions are simulated by considering a single polymer chain suspended in a bulk Newtonian fluid. The polymer is modeled considering a fully flexible chain consisting of a sequence of $N_b$ beads connected by $(N_b-1)$ entropic springs~\cite{Bird1987vol2,Jain_PRL2012,KumarPrkash2003,Sunthar2006}. The force along the connector vector between a bead pair is modeled by a finitely extensible non-linear elastic (FENE) spring potential, given by $U_s(Q) = -\frac{1}{2}HQ_0^2\ln(1-Q^2/Q_0^2)$, with a maximum non-dimensional extensibility of the spring, $Q_0/l_H=7.071$ (which is equivalent to FENE $b$-parameter of $50$)~\cite{Santra2022}. Such a large value of FENE $b$-parameter may lead to self-crossing of the chains which is an important issue to consider in the study of dynamics of polymer solutions in highly concentrated entangled regime. However, since in this study we are mostly interested in the dilute unentangled regime, such a value of FENE $b$-parameter will not qualitatively affect the computed properties. Note that all the length and time scales in the simulations are non-dimensionalized with $l_H=\sqrt{k_BT/H}$ and $\lambda_H = \zeta/4H$, respectively, where $k_B$ is the Boltzmann constant, $H$ is the spring constant and $\zeta = 6\pi\eta_s a$ is the coefficient of Stokes friction for a bead of radius $a$ and solvent viscosity $\eta_s$. The solvent quality parameter $z$ is controlled by the pair-wise excluded volume (EV) interaction modeled by a narrow Gaussian potential~\cite{Ottinger1996,Ravi1999,KumarPrkash2003}, given by the following non-dimensional expression,
\begin{equation}
    E(\bm{r^*}_{\nu\mu}) = \frac{z^*}{d^{*3}}\exp\left(-\frac{1}{2}\frac{\bm{r}^{*2}_{\nu\mu}}{d^{*2}}\right),
\end{equation}
where, $\bm{r^*}_{\nu\mu}$ is the distance vector between bead $\nu$ and $\mu$, $z^*$ is a non-dimensional parameter defining the strength of EV interaction and the non-dimensional parameter $d^*$ specify the range of EV interaction. Choosing the narrow Gaussian potential for modeling EV interaction provides the flexibility to smoothly change the solvent quality between $\theta$ and athermal limits by altering the value of $z^*$.  Accordingly, the solvent quality parameter for a bead-spring chain model with narrow Gaussian potential is defined by $z=z^*\sqrt{N_b}$~\cite{KumarPrkash2003,Sunthar2006,Santra2022}. The tangential active force acting on bead $\mu$ is modeled by the following non-dimensional equation~\cite{Panda_PRE2025,Panda_Macromol2025},
\begin{equation}
\mathbf{F}_\mu^{a} = \frac{f_a}{(k_BT/l_H)} \left( \frac{\hat{\mathbf{t}}_{\mu+1} + \hat{\mathbf{t}}_{\mu}}{2}\right),\,\,\text{for}\,\,\mu \in \{2,3,.., N_b-1\}
\end{equation}
where, $\hat{\mathbf{t}}_\mu = (\mathbf{r^*}_\mu - \mathbf{r^*}_{\mu-1})/(|\mathbf{r^*}_\mu - \mathbf{r^*}_{\mu-1}|)$, and $Pe=(f_a l_H/k_B T)$ is the P{\'e}clet number determining the strength of activity. For the terminal beads, $\mu=1$ and $\mu=N_b$, the active force is defined as $\mathbf{F}_1^{a}=Pe(\hat{\mathbf{t}}_{2}/2)$ and $\mathbf{F}_{N_b}^{a}=Pe(\hat{\mathbf{t}}_{N_b}/2)$, respectively. 

Time evolution of the polymer chain configuration is computed by solving the following non-dimensional It{\^o} stochastic differential equation for the bead position vector, $\mathbf{R}$, obtained by incorporating the spring potential, EV interaction and tangential active force.
\begin{equation}\label{eq:sm1}
    d\mathbf{R} = (\mathbf{K}\cdot\mathbf{R} + \frac{1}{4}\mathbf{D}\cdot\mathbf{F}^{\phi})d t^* + \frac{1}{\sqrt{2}}\mathbf{B}\cdot{}d\mathbf{W}
\end{equation}
Here, $\mathbf{R}$ consists of the coordinates of the position vectors of $N_b$ beads. $\mathbf{K}$ is a block matrix consisting of $N_b\times N_b$ blocks of $3\times 3$ matrices, accounting for the deformation due to an imposed flow field. Since the present study is carried out under equilibrium conditions, $\mathbf{K}$ is considered as a zero matrix. $\mathbf{F}^{\phi} = \mathbf{F}^{s} + \mathbf{F}^{e} + \mathbf{F}^a$ is the total non-hydrodynamic force acting on a bead consisting of contributions from spring potential ($\mathbf{F}^s$), excluded volume interaction ($\mathbf{F}^e$) and tangential activity ($\mathbf{F}^a$). The $\mu\nu$-th component of the diffusion tensor $\mathbf{D}$ (which is also a block matrix of $N_b\times N_b$ blocks with dimensions of $3\times 3$) is defined as $\mathbf{D}_{\mu\nu} = \delta_{\mu\nu}\bm{\delta}+\bm{\Omega}_{\mu\nu}$, where $\delta_{\mu\nu}$ is the Kronecker delta function, $\bm{\delta}$ is the unit tensor and $\bm{\Omega}_{\mu\nu}$ is the hydrodynamic interaction (HI) tensor for bead pair $\mu$, $\nu$. It is to be noted that in this study we have only computed equilibrium static properties, such as, mean squared radius of gyration and swelling ratio, which are independent of hydrodynamic interactions~\cite{Bird1987vol2,Doi1988theory}. Therefore, we have excluded HI from our model equations to reduce computational complexity. In general, $\bm{\Omega}_{\mu\nu}$ is modeled by regularization techniques such as by using Rotne-Prager-Yamakawa tensor~\cite{Rotne1969,Yamakawa1970}. However, since in the present study we have neglected HI, $\mathbf{\Omega}_{\mu\nu}$ is set to zero. $d\mathbf{W}$ represents a $3 N_b$-dimensional Wiener process. Matrix $\mathbf{B}$ is obtained by decomposition of the diffusion tensor following the rule $\mathbf{D}=\mathbf{B}\cdot\mathbf{B}^{T}$. 

\subsection{Semi-implicit predictor-corrector scheme}

The governing equation, Eq.~(\ref{eq:sm1}), is discretized and solved numerically by using a semi-implicit predictor-corrector scheme proposed in earlier studies for passive polymer solutions by \citet{Kroger2000,Somasi2002} and, \citet{Prabhakar2004}. Here, we have extended the method for active polymer chains. The scheme involves a two-step solution procedure, a \emph{predictor step} followed by a \emph{corrector step}. 

The \emph{predictor step} involves discretization of the governing differential equation using an explicit Euler-Maruyama scheme which gives the following configuration of the polymer chain, $\mathbf{\tilde{R}}_{n+1}$, at the ($n+1$)$^{th}$ time step depending on their values at the $n^{th}$ time step.

\begin{widetext}
\begin{equation}
 \mathbf{\tilde{R}}_{n+1} = \mathbf{R}_n + \left[\frac{1}{4}\mathbf{D}_n\cdot\mathbf{F}_n^s+\frac{1}{4}\mathbf{D}_n\cdot\mathbf{F}_n^e+\frac{1}{4}\mathbf{D}_n\cdot\mathbf{F}_n^a\right]\Delta t^* + \frac{1}{\sqrt{2}}\mathbf{B}_n\cdot\Delta\mathbf{W}_n  
\end{equation}
\end{widetext}

\noindent Here, $\Delta t^*$ is the non-dimensional time step size and the Wiener increment $\Delta\mathbf{W}_n$ is sampled from a real-valued Gaussian distribution with mean zero and variance $\Delta t^*$. The bead position coordinates estimated from the predictor step is updated in the \emph{corrector} step by the following equation.

\begin{widetext}
\begin{eqnarray}\label{eq:corr}
    \mathbf{R}_{n+1} &=& \mathbf{R}_n + [\frac{1}{8}\mathbf{D}_n\cdot(\mathbf{F}_n^s+\mathbf{F}_{n+1}^s) \nonumber \\
   & & +\frac{1}{8}\mathbf{D}_n\cdot(\mathbf{F}_n^e+\mathbf{\tilde{F}}_{n+1}^e) + \frac{1}{8}\mathbf{D}_n\cdot(\mathbf{F}_n^a + \mathbf{F}_{n+1}^a)]\Delta t^* + \frac{1}{\sqrt{2}}\mathbf{B}_n\cdot\Delta\mathbf{W}_n 
\end{eqnarray}
\end{widetext}

\noindent In the above equation, all the quantities, except the FENE spring force $\mathbf{F}^s$ and the active force $\mathbf{F}^a$, are evaluated based on $\mathbf{R}_n$ or the position vector $\mathbf{\tilde{R}}_{n+1}$, estimated in the predictor step. $\mathbf{F}^s_{n+1}$ and $\mathbf{F}^a_{n+1}$ are forces evaluated in terms of $\mathbf{R}_{n+1}$, i.e. $\mathbf{F}^s$ and $\mathbf{F}^a$ are treated implicitly in our calculations. The corrected value of the position vector for a bead $\nu$ is obtained by expressing Eq.~(\ref{eq:corr}) in terms of the connector vector $\bm{Q}_{\nu}^* = \bm{r}_{\nu+1}^*-\bm{r}_{\nu}^*$, by introducing an operator $\mathcal{D}_{\nu}$ and subsequently solving the resultant system of non-linear algebraic equations. The operator $\mathcal{D}_{\nu}$ can be operated on any $3N\times M$ matrix to generate a $3\times M$ matrix with elements obtained by subtraction of the three rows corresponding to the $\nu$th block of rows from the $(\nu+1)$th block. This operation results in the following expression for the connector vector $\bm{Q}_{\nu}^*$,

\begin{eqnarray}\label{eq:Qnu}
    \bm{Q}_{\nu,n+1}^* &=& \mathcal{D}_{\nu}[\mathbf{\tilde{Y}}_{n+1}] + \frac{1}{8}\mathcal{D}_{\nu}[\mathbf{D}_n\cdot\mathbf{F}_{n+1}^s]\,\Delta t^* \nonumber \\
    & & + \frac{1}{8}\mathcal{D}_{\nu}[\mathbf{D}_n\cdot\mathbf{F}_{n+1}^a]\,\Delta t^*
\end{eqnarray}
\noindent where, 
\begin{eqnarray}
\mathbf{\tilde{Y}}_{n+1} &=& \mathbf{R}_n + [\frac{1}{8}\mathbf{D}_n\cdot\mathbf{F}_n^s +\frac{1}{8}\mathbf{D}_n\cdot(\mathbf{F}_n^e+\mathbf{\tilde{F}}_{n+1}^e) \nonumber \\
   & &  + \frac{1}{8}\mathbf{D}_n\cdot\mathbf{F}_n^a]\,\Delta t^* + \frac{1}{\sqrt{2}}\mathbf{B}_n\cdot\Delta\mathbf{W}_n
\end{eqnarray}

\noindent Finally, we set up a non-linear equation of the following form by adding $(1/4)\mathbf{F}_{\nu}^{c*}\,\Delta t^*$ on both the sides of Eq.~(\ref{eq:Qnu}), considering $\mathbf{F}_{\nu}^{c*} = \mathbf{F}_{\nu+1}^{s*}-\mathbf{F}_{\nu}^{s*}$.

\begin{equation}\label{eq:Q_itr}
    \bm{Q}_{\nu,n+1}^{*(i)} + \frac{\Delta t^*}{4}\mathbf{F}_{\nu}^{c*(i)} = \bm{\Gamma}_{\nu,n+1}^{(i-1)}
\end{equation}
\noindent where,
\begin{eqnarray}
    \bm{\Gamma}_{\nu,n+1}^{(i-1)} &=& \mathcal{D}_{\nu}\,[\mathbf{\tilde{Y}}_{n+1}] + \frac{1}{8}\mathcal{D}_{\nu}\,[\mathbf{D}_n\cdot\mathbf{F}_{n+1}^{s(i-1)}]\,\Delta t^* \nonumber \\
    & &  + \frac{1}{8}\mathcal{D}_{\nu}\,[\mathbf{D}_n\cdot\mathbf{F}_{n+1}^{a(i-1)}]\,\Delta t^* + \frac{\Delta t^*}{4}\mathbf{F}_{\nu}^{c*(i-1)}
\end{eqnarray}

\noindent Here, all the quantities on the left hand side of Eq.~(\ref{eq:Q_itr}) are evaluated in the $i^{th}$ iteration step and the terms on the right hand side are evaluated based on the values calculated in the $(i-1)^{th}$ iteration. In general, the iterations are continued until $|\mathbf{R}_{n+1}^{(i)}-\mathbf{R}_{n+1}^{(i-1)}|/|\mathbf{R}_{n+1}^{(i)}| < \epsilon$, where $\epsilon$ is a specified tolerance value. However, in the present study we replace $\mathbf{F}_{\nu}^{c*(i)}$ by the FENE spring force law, to obtain the following equation.

\begin{equation}
    \left(1+\frac{1}{4}\frac{\Delta t^*}{1-|\bm{Q}^{*(i)}_{\nu,n+1}|^2/b}\right)\bm{Q}_{\nu,n+1}^{*(i)} = \bm{\Gamma}_{\nu,n+1}^{(i-1)}\,,
\end{equation}

\noindent where, $\sqrt{b}$ is the maximum extensibility of the FENE spring. Substituting $x= |\bm{Q}_{\nu,n+1}^{*(i)}|$ and $\mathcal{G} = |\bm{\Gamma}_{\nu,n+1}^{(i-1)}|$ in the above equation results into the following cubic equation.

\begin{equation}\label{eq:cubic}
    x^3 -\mathcal{G}x^2-b\left(1+\frac{\Delta t^*}{4}\right)x + b\,\mathcal{G} = 0
\end{equation}

\noindent Eq.~(\ref{eq:cubic}) can be solved analytically to obtain exactly one root in the domain between $0$ and $\sqrt{b}$, which will give the corrected value of the connector vector. This algorithm provides unconditional stability by keeping the magnitudes of all the connector vectors within the domain $(0,\sqrt{b})$. Therefore, we could use a much larger time step size for the BD simulations resulting in a reduction of the overall CPU time.

\begin{figure*}[ptbh]
  \centerline{
 \resizebox{\textwidth}{!}{ \begin{tabular}{cc}
        \includegraphics[width=14cm,height=!]{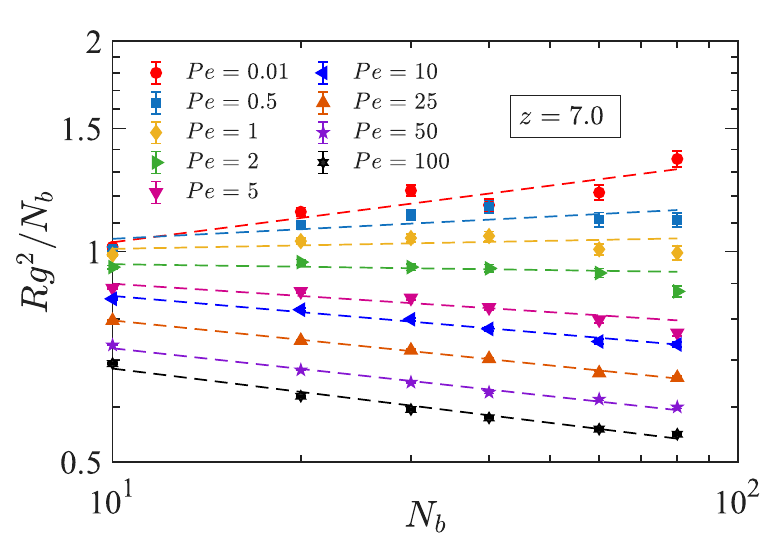} 
    & 
       \includegraphics[width=13cm,height=!]{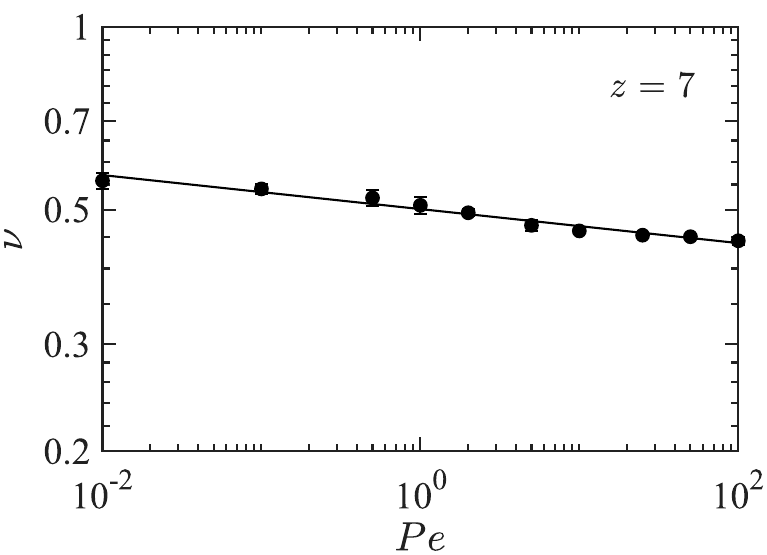} \\[5pt]
           {\huge{(a)}}  & 
       {\huge{(b)}} \\
       \multicolumn{2}{c}{\includegraphics[width=14cm,height=!]{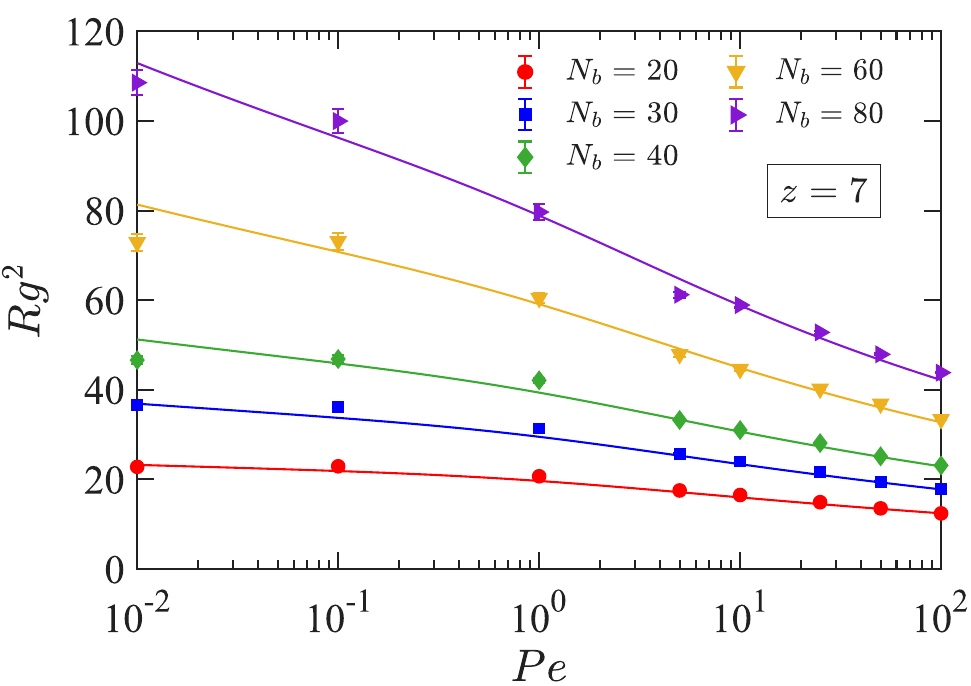} }\\[5pt]
       \multicolumn{2}{c}{\huge{(c)}}     \\
  \end{tabular} }}
\caption{\small{(a) Scaling of $R_g^2/N_b$ as a function of number of beads $N_b$ for a range of values of $Pe$ at solvent quality parameter $z=7$. The dotted lines represent power-law fit to the simulation data. (b) Variation of exponent $\nu$ with $Pe$ for solvent quality parameter $z=7$. The solid line is a power-law fit to the data given by $\nu=0.5011\,Pe^{-0.028}$. (c) Variation of square of the gyration radius as a function of P{\'e}clet number at different values of chain length, $N_b$, for $z=7$. The solid lines represent a fit to the simulation data.}} 
\label{fig:Rg2_scale}
\end{figure*}

In the present simulations, we have used $\Delta t^*=0.001$ as the non-dimensional time step size. Mean values of the properties are calculated by a block ensemble average over 192 independent trajectories, where each trajectory consists of data collected over 1000 equispaced time instances. Each trajectory is equilibrated for five non-dimensional Rouse relaxation time ($\tau^*_R$), followed by a production run over $5\,\tau^*_R$, where the non-dimensional Rouse relaxation time for a linear chain of $N_b$ beads is defined as, $\tau_R^* = \displaystyle\frac{1}{2\,\sin^2(\pi/2N_b)}$.

%%%%%%%%%%%%%%%%%%%%%%%%%%%%%%%%%%%%
\section{\label{sec:results}Results and discussion}
%%%%%%%%%%%%%%%%%%%%%%%%%%%%%%%%%%%%%
\begin{figure*}[ptbh]
  \centerline{
 \resizebox{\textwidth}{!}{ \begin{tabular}{cc}
        \includegraphics[width=14cm,height=!]{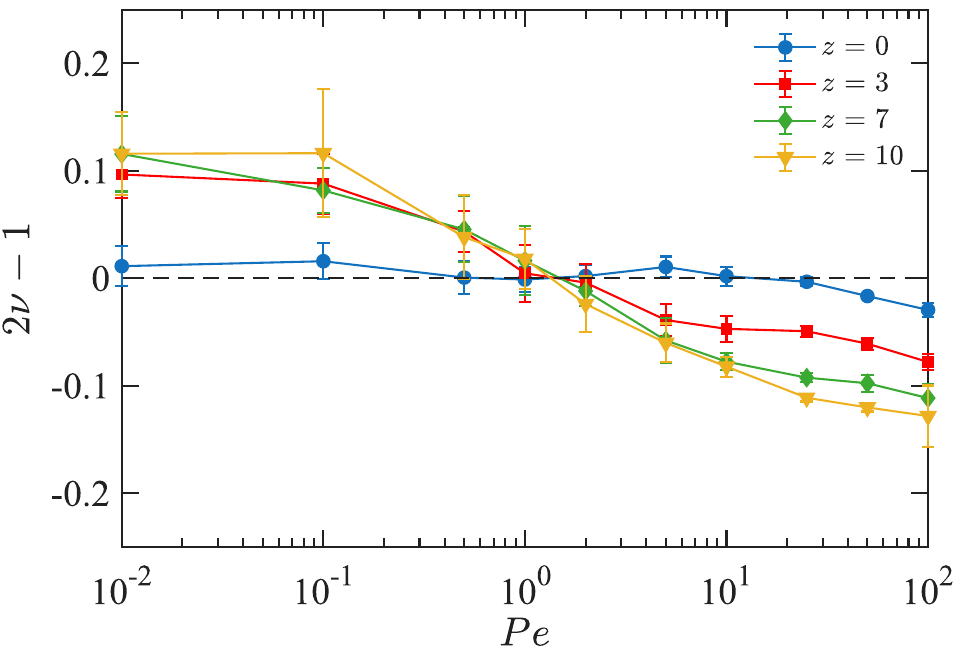} 
    & 
       \includegraphics[width=13.5cm,height=!]{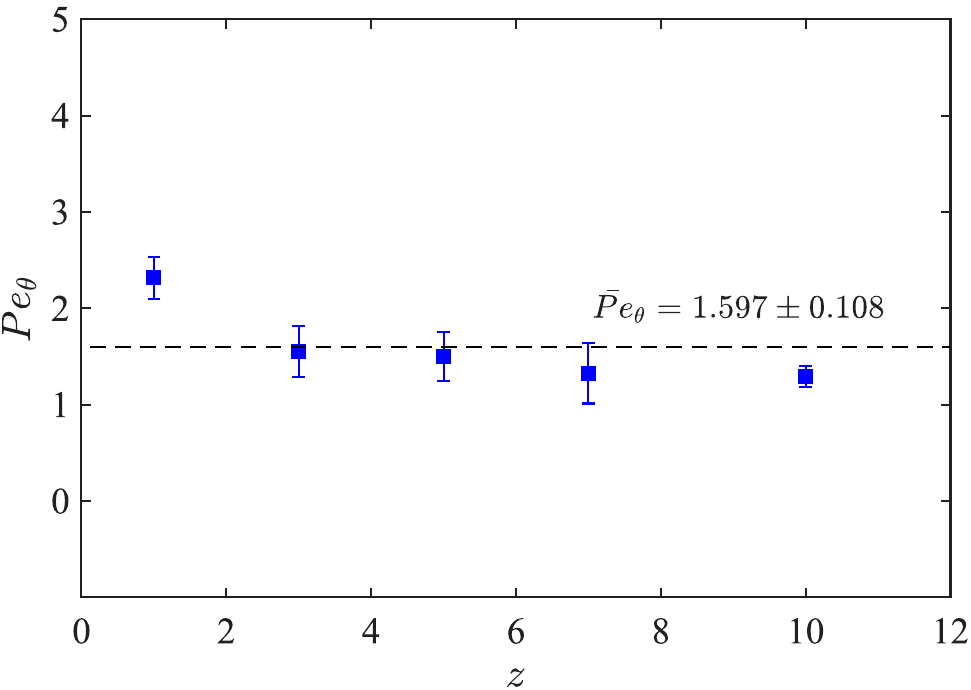} \\[5pt]
           {\huge{(a)}}  & 
       {\huge{(b)}} \\
  \end{tabular} }}
\caption{\small{(a) Variation of exponent $\nu$, expressed in the form $(2\nu-1)$, with P{\'e}clet no. $Pe$ at different values of solvent quality parameter $z$. (b) Variation of P{\'e}clet no. at the $\theta$ condition, $Pe_{\theta}$, as a function of $z$.}} 
\label{fig:Pe_theta}
\end{figure*}

Simulations are carried out at different values of $Pe$, over a range of values of number of beads ($N_b$) for $z=0$ to $10$ with a constant value of $d^*=1$. The equilibrated values of mean squared radius of gyration, $R_g^2\equiv\langle R_g^2\rangle$, is evaluated, where $\langle R_g^2\rangle$ denotes a block ensemble average calculated over $192$ independent trajectories. It is to be noted that for a given $z$, the quantity $z^*$ is adjusted  with $N_b$ such that $z=z^*\sqrt{N_b}$ remains unchanged. At any value of $z>0$, the variation of $R_g^2$ with $N_b$ indicates that $R_g\sim N_b^{\nu}$, where the exponent $\nu$ decreases monotonically with P{\'e}clet number following the relation $\nu=A\, Pe^{-\beta}$. This is illustrated in Fig.~\ref{fig:Rg2_scale} (a) and \ref{fig:Rg2_scale} (b) for $z=7$ with values of $Pe$ ranging from $0.01$ to $100$. Fig.~\ref{fig:Rg2_scale} (a) and \ref{fig:Rg2_scale} (b) further demonstrate that tangential activity results in a coil-to-globulelike transition of active polymer chains in the thermal crossover regime, similar to that observed in some earlier work~\cite{Bianco2018,Jain_Marco2022} for active polymers with EV interaction. Moreover, the variation of $R_g^2$ with $Pe$, as shown in Fig.~\ref{fig:Rg2_scale}~(c), can be fitted by the following functional form as proposed by \citet{Bianco2018} for self-avoiding and Gaussian chains.
\begin{equation}
    R_g^2 = b^2\left(\frac{a_g+h_g\ln(Pe)}{(1+Pe)^{c_g}}\right)^2\,N_b^{2\nu}
\end{equation}
Here, $b$ denotes the size of a monomer unit and the fitting parameters $a_g$, $h_g$ and $c_g$ are independent of $N_b$, $Pe$ and $z$.  The term $a_g+h_g\ln(Pe)$ predicts the variation of $R_g^2$ with $Pe$ in the limit of low P{\'e}clet number and $1/(1+Pe)^{c_g}$ captures the decreasing trend of $R_g^2$ at larger $Pe$.

As presented in Fig.~\ref{fig:Rg2_scale}~(b), for any tangentially active polymer chain in a dilute solution at $z>0$, the decay in the swelling exponent $\nu$ with $Pe$ indicates that the effective solvent quality undergoes a transition from a good solvent condition ($\nu \approx 0.55$) to poor solvent ($\nu < 0.5$). This suggests that for a given value of $z$ we can extract a value of P{\'e}clet number at which the effective solvent quality replicates $\theta$-condition (i.e. $\nu=0.5$). Accordingly, in Fig.~\ref{fig:Pe_theta}~(a), the variation of the quantity $(2\nu-1)$ as a function of $Pe$ is presented for different values of solvent quality parameter $z$ and the value of P{\'e}clet number at which $2\nu-1=0$ (denoted by $Pe_{\theta}$) is estimated in each case. Notably, as presented in Fig.~\ref{fig:Pe_theta}~(b), $Pe_{\theta}$ values are found to be independent of the solvent quality parameter $z$. This implies that there is a unique value of P{\'e}clet number $\bar{Pe}_{\theta}\approx1.597\pm 0.108$ at which the effective solvent quality in a tangentially active polymer solution becomes $\theta$, irrespective of the strength of EV interaction. Considering that $\bar{Pe}_{\theta}$ produces a $\theta$-condition at any given $z$, we have defined a swelling ratio ($\alpha_g^2$) in terms of the mean squared radius of gyration, given by $\alpha_g^2=R_g^2(z,Pe)/R_{g\theta}^2$, where, $R_g^2(z,Pe)$ denotes the mean squared radius of gyration at a given $z$ and $Pe$ ($<\bar{Pe}_{\theta}$), and $R_{g\theta}^2$ represents the value of $R_g^2$ at $z=0$.

For a wide range of polymer solvent combination, the swelling ratio for the gyration radius of passive linear polymers relative to their $\theta$ state is found to follow a universal curve with respect to the solvent quality parameter $z$~\cite{Miyaki1981,Santra2019,KumarPrkash2003,Sunthar2006}. This universal curve can be represented in a parametrize form given by the following expression obtained from renormalize group calculations~\cite{schafer1999,KumarPrkash2003}, 
\begin{equation}
\tilde{f} (z) = (1 + a z + b z^2 + c z^3)^m,
\label{eqn:swellratio}
\end{equation}
\noindent where, $\tilde{f}(z)=\alpha_g^2$, and $a\approx9.5286$, $b=19.48 \pm 1.28$, $c=14.92 \pm 0.93$ and $m=0.1339 \pm 0.0006$ are the estimated values of the parameters~\cite{KumarPrkash2003,Santra2022}. It is important to note that this universal relation can be directly used to determine the solvent quality of any passive polymer-solvent system at a given temperature and polymer molecular weight, $M_w$, just by solving Eq.~(\ref{eqn:swellratio}) for unknown the $z$, provided $\alpha_g^2$ is known. While for passive polymer solutions the solvent quality is solely dependent on the parameter $z$, incorporation of activity introduces an additional parameter $Pe$ which can alter the effective solvent quality. Furthermore, the activity affects the universal relationship between the swelling ratio and $z$. The effects of tangential activity on the universal swelling behaviour are discussed in the subsequent paragraphs. 

\begin{figure*}[ptbh]
  \centerline{
 \resizebox{\textwidth}{!}{ \begin{tabular}{cc}
        \includegraphics[width=14cm,height=!]{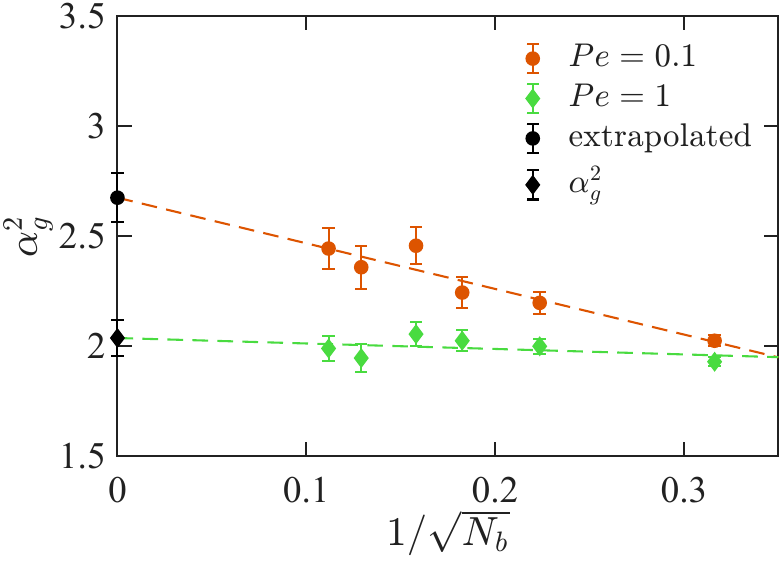} 
    & 
       \includegraphics[width=14cm,height=!]{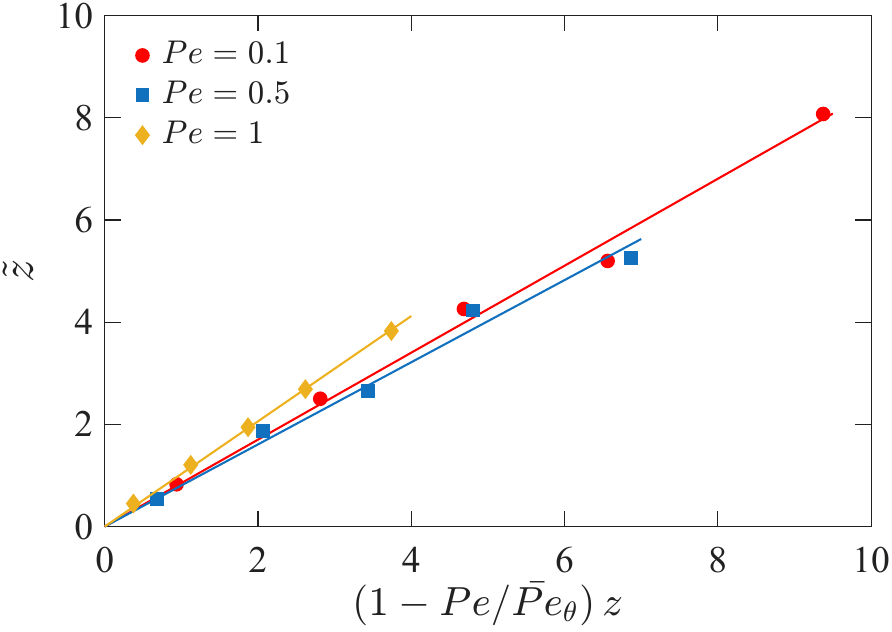} \\[5pt]
           {\huge{(a)}}  & 
       {\huge{(b)}} \\
       \multicolumn{2}{c}{\includegraphics[width=14cm,height=!]{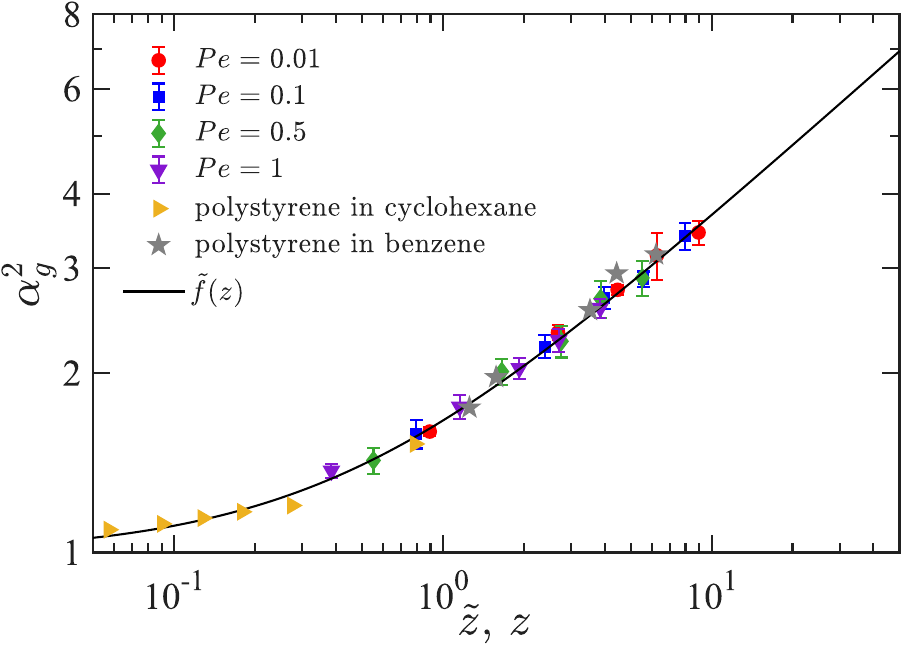} }\\[5pt]
       \multicolumn{2}{c}{\huge{(c)}}     \\
  \end{tabular} }}
\caption{\small{(a) Plot of $\alpha_g^2$ versus $1/\sqrt{N_b}$ to determine the swelling ratio in the limit of infinite chain length  at $z=5$ for $Pe=0.1$ and $Pe=1$, using successive fine-graining. (b) Variation of renormalized solvent quality parameter $\tilde{z}$ for tangentially active polymer solutions with $z$ and $Pe$. (c) Variation of swelling ratio $\alpha_g^2$ as a function of renormalized solvent quality parameter $\tilde{z}$ for active polymers and its comparison with the swelling ratio for passive polymer solutions. The solid line represents the function $\tilde{f}(z)$ defined in Eq.~(\ref{eqn:swellratio}). The data for the solution of polystyrene in cyclohexane and benzene are reproduced from the experiments by \citet{Miyaki1981}.}} 
\label{fig:uni_swell}
\end{figure*}

The universal form of the swelling ratio presented in Eq.~(\ref{eqn:swellratio}) is typically applicable in the limit long chain length where the ratio becomes independent of the model parameters~\cite{KumarPrkash2003,Sunthar2006}. However, in the present study the simulations are carried out with finite number of beads. Therefore, to evaluate the swelling ratio in the long chain limit we have used a successive fine-graining technique proposed by \citet{Prabhakar_Jor2004}. In this method, we simulate active polymer chains by progressively increasing the number of beads starting from $N_b=10$ and calculate $\alpha_g^2$ ratio for each $N_b$. Finally, the values of the ratio at different number of beads are plotted against $1/\sqrt{N_b}$, and $\alpha_g^2$ is estimated in the limit of infinite chain length by extrapolating the data to $N_b\rightarrow \infty$ (i.e. $1/\sqrt{N_b}\rightarrow 0$). An illustration of this method is provided in Fig.~\ref{fig:uni_swell}~(a), where, $\alpha_g^2$ in the limit of $N_b\rightarrow\infty$ is evaluated for $Pe=0.1$ and $1$. It is to be noted that for various material properties in polymer solutions, the leading order correction is of order $1/\sqrt{N_b}$~\cite{RaviPrakash2001,schafer1999}, which becomes the rationale for choosing $1/\sqrt{N_b}$ as the abscissa in Fig.~\ref{fig:uni_swell}~(a). Using the extrapolated values of $\alpha_g^2$, an effective solvent quality parameter ($\tilde{z}$) for the active polymer solutions is estimated from Eq.~(\ref{eqn:swellratio}) by solving for $z=\tilde{z}$, equating $\tilde{f}(\tilde{z})$ to $\alpha_g^2$. As shown in Fig.~\ref{fig:uni_swell} (b), for different values of P{\'e}clet number, $\tilde{z}$ is found to vary linearly with the solvent quality parameter $z$, such that $\tilde{z}=\hat{k}(1-Pe/\bar{Pe}_{\theta})\,z$, where $\hat{k}$ is an activity-dependent constant pre-factor. The quantity $(1-Pe/\bar{Pe}_{\theta})\,z$ is a renormalized solvent quality parameter for the active polymer solutions which depends on $T$, $M_w$ and $Pe$. Notably, as presented in Fig.~\ref{fig:uni_swell} (c), at different values of $Pe$, the variation of swelling ratio ($\alpha_g^2$) for tangentially active polymers as a function of the effective solvent quality parameter $\tilde{z}$ follows the same universal curve, given by Eq.~(\ref{eqn:swellratio}), as that of the passive polymers. Furthermore, we have included experimental data for solutions of polystyrene in cyclohexane and benzene by \citet{Miyaki1981} in Fig.~\ref{fig:uni_swell} (c) to show that the swelling ratio for different polymer-solvent combinations can also be described by the same curve. 

Moreover, the results suggest that tangential activity induces deswelling of the polymer chains, leading to a coil-to-globulelike transition, which can be explained as a consequence of the reduction in magnitude of the solvent quality parameter by a factor of $(1-Pe/\bar{Pe}_{\theta})$. This leads to a novel scaling relation between the thermal blob size of tangentially active polymer and P{\'e}clet number, as presented in the following section.

%%%%%%%%%%%%%%%%%%%%%%%%%%%%%%%%%%%%%%%%%%%%%%%%%%%%%%%%%%%%%%%%%%%%%%%%%
\subsection{Scaling of thermal blobs}
%%%%%%%%%%%%%%%%%%%%%%%%%%%%%%%%%%%%%%%%%%%%%%%%%%%%%%%%%%%%%%%%%%%%%%%%%
\begin{figure}[htbp]
    \centering
    \includegraphics[width=0.6\linewidth]{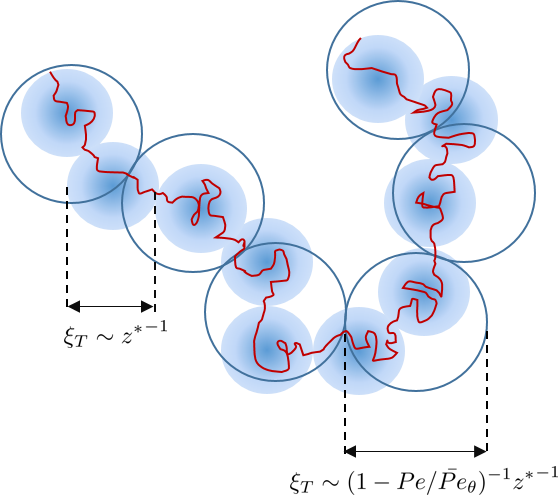}
    \vspace{0.1in}
    \caption{\small{Schematic of a polymer chain represented as a sequence of thermal blobs, depicted by solid or open blue-colored circles. Smaller solid circles represent the size of a thermal blob ($\xi_T$) in absence of activity, whereas, the larger open circles represent $\xi_T$ for the polymer chain with tangential activity.}}
    \label{fig:blob_pic}
\end{figure}

\noindent Within the framework of polymer blob scaling ansatz, for dilute polymer solutions, the number of thermal blobs $\mathcal{N}_T$ on a polymer chain is only dependent of the solvent quality parameter $z$ as, $\mathcal{N}_T=z^2$~\cite{Jain_PRL2012,degennes1979,Grosberg1994,Santra2021}. Thus, the number of monomers in a thermal blob ($g_T$) can be defined as, $g_T = N_b/\mathcal{N}_T\sim \hat{\tau}^{-2}$, where $N_b$ is the total number of monomer units on a chain (which is proportional to $M_w$) and $\hat{\tau}(T)$ is the temperature dependent term in the solvent quality parameter. It is to be noted that for a narrow Gaussian potential, $g_T\sim {z^*}^{-2}$. Therefore, the thermal blob size is given by, $\xi_T=b\,g_T^{1/2}\sim {1/z^*}$, where $b$ is the size of a monomer unit. Implementing this analysis for tangentially active polymer chains results into the following scaling relation for the thermal blob size,
\begin{equation}
\xi_T\sim (1-Pe/\bar{Pe}_{\theta})^{-1}{z^*}^{-1},
\end{equation}
which shows that $\xi_T$ grows with P{\'e}clet number by a factor of $1/(1-Pe/\bar{Pe}_{\theta})$, leading to a transition of the effective solvent quality from an athermal condition towards $\theta$ limit. This is illustrated by a schematic diagram in Fig.~\ref{fig:blob_pic}, where a polymer chain is represented as a sequence of thermal blobs (indicated by solid or open blue circles), whose size grows with increasing $Pe$. The coil-to-globule transition in the chain conformation is further analysed by computing the variation of shape anisotropy and asphericity of the polymer chains with activity, which is discussed in the following section.

%%%%%%%%%%%%%%%%%%%%%%%%%%%%%%%%%%%%%%%%%%%%%%%%%%%%%%%%%%%%%%%%%%%%%%%%%
\subsection{Shape anisotropy and asphericity of the polymer conformation}
%%%%%%%%%%%%%%%%%%%%%%%%%%%%%%%%%%%%%%%%%%%%%%%%%%%%%%%%%%%%%%%%%%%%%%%%%

\begin{figure*}[ptbh]
  \centerline{
 \resizebox{\textwidth}{!}{ \begin{tabular}{cc}
        \includegraphics[width=16cm,height=!]{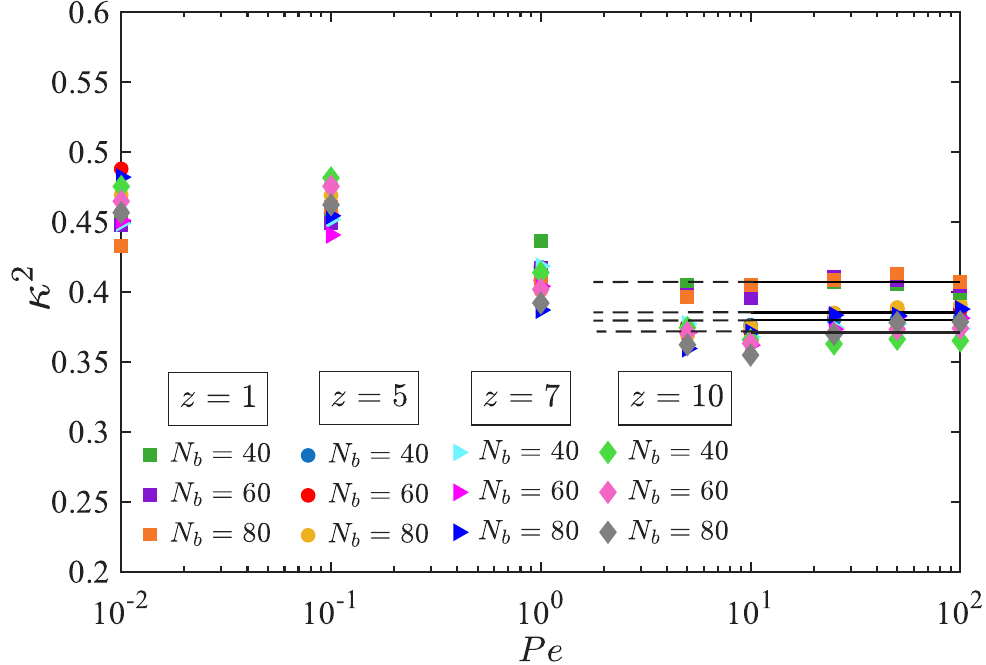} 
       &
       \includegraphics[width=15cm,height=!]{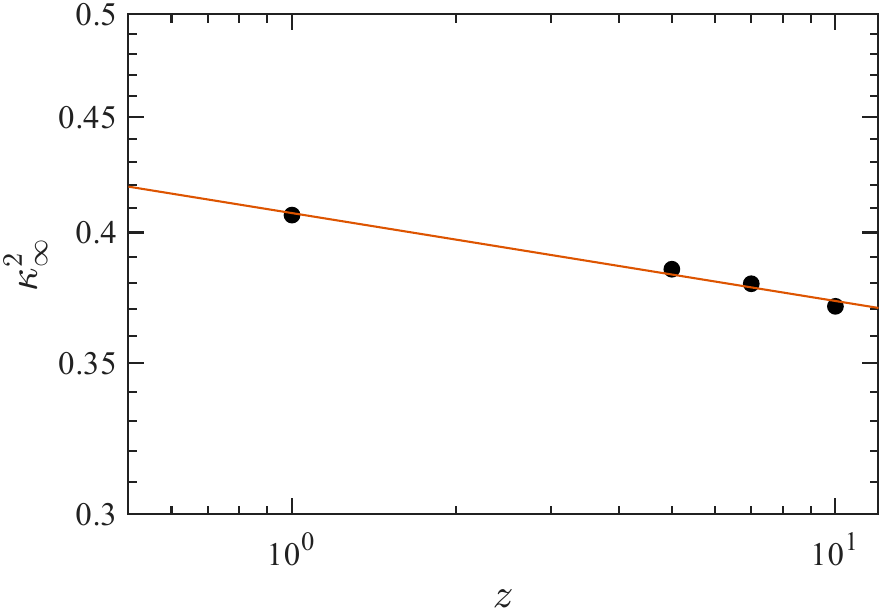} \\[5pt]
        \huge{(a)}  & 
       \huge{(b)} \\
  \end{tabular} }}
\caption{\small{(a) Variation of shape anisotropy $\kappa^2$ with P{\'e}clet number $Pe$ for different values of number of beads per chain $N_b$, at $z=1,\,5,\, 7$ and $10$. The dashed-solid horizontal lines represent the asymptotic values of $\kappa^2$ in the limit of large $Pe$. The error bars are smaller than the symbol size. (b) Variation of the asymptotic value of shape anisotropy, $\kappa^2_{\infty}$, as a function of $z$. The solid line represents a power law fit to the data given by, $\kappa^2_{\infty} = 0.408\,z^{-0.039}$.}} 
\label{fig:shape_aniso}
\end{figure*}

\noindent As discussed earlier, increasing the strength of tangential activity leads to deswelling of the polymer chains in the thermal crossover regime. We have analyzed this configurational transformation by computing the shape descriptors derived from the invariants of the gyration tensor. The gyration tensor ($S$) is computed from the instantaneous polymer chain configurations using the following definition~\cite{Arkin2013},

\begin{widetext}
\begin{equation}
S=\frac{1}{N_b}\begin{pmatrix}
  \sum\limits_{i=1}^{N_b}(x_i-x_\text{cm})^2 & \sum\limits_{i=1}^{N_b}(x_i-x_\text{cm})(y_i-y_\text{cm}) &  \sum\limits_{i=1}^{N_b}(x_i-x_\text{cm})(z_i-z_\text{cm})\\
   \sum\limits_{i=1}^{N_b}(x_i-x_\text{cm})(y_i-y_\text{cm}) & \sum\limits_{i=1}^{N_b}(y_i-y_\text{cm})^2 & \sum\limits_{i=1}^{N_b}(y_i-y_\text{cm})(z_i-z_\text{cm})\\
   \sum\limits_{i=1}^{N_b}(x_i-x_\text{cm})(z_i-z_\text{cm}) & \sum\limits_{i=1}^{N_b}(y_i-y_\text{cm})(z_i-z_\text{cm}) &  \sum\limits_{i=1}^{N_b}(z_i-z_\text{cm})^2
\end{pmatrix},	
\end{equation}
\end{widetext}

\noindent where, $N_b$ is the number of beads on a chain, $(x_i,y_i,z_i)$ represents the position coordinates of bead $i$, and $(x_\text{cm},y_\text{cm},z_\text{cm})$ denotes the position coordinates of the center of mass. Three eigenvalues of the gyration tensor $\lambda_1,\,\lambda_2$ and $\lambda_3$, of the order $\lambda_1\le \lambda_2\le \lambda_3$, are calculated and used to evaluate the invariants. The first invariant of $S$ gives the value of radius of gyration squared ($R_g^2$), defined as follows,
\begin{equation}
    \text{Tr}\,S = \lambda_1 + \lambda_2 + \lambda_3 = R_g^2 \, ,
\end{equation}
where, Tr$\, S$ is the trace of matrix $S$. The shape anisotropy ($\kappa^2$) of the polymer chain configuration is computed from the second invariant by using the following definition~\cite{Arkin2013}.
\begin{equation}
  \kappa^2 = 1 -3 \frac{\lambda_1\lambda_2+\lambda_2\lambda_3+\lambda_3\lambda_1}{(\lambda_1+\lambda_2+\lambda_3)^2}   
\end{equation}
$\kappa^2$ takes values between $0$ and $1$, where, $\kappa^2=0$ indicates a completely symmetric configurational structure and $\kappa^2=1$ denotes an ideal linear chain. Finally, the deviation from spherical symmetry is characterised by the asphericity parameter $b_{sp}$, defined as, $b_{sp}=\lambda_1-\displaystyle\frac{1}{2}(\lambda_2+\lambda_3)$.

\begin{figure*}[ptbh]
  \centerline{
 \resizebox{\textwidth}{!}{ \begin{tabular}{cc}
        \includegraphics[width=14cm,height=!]{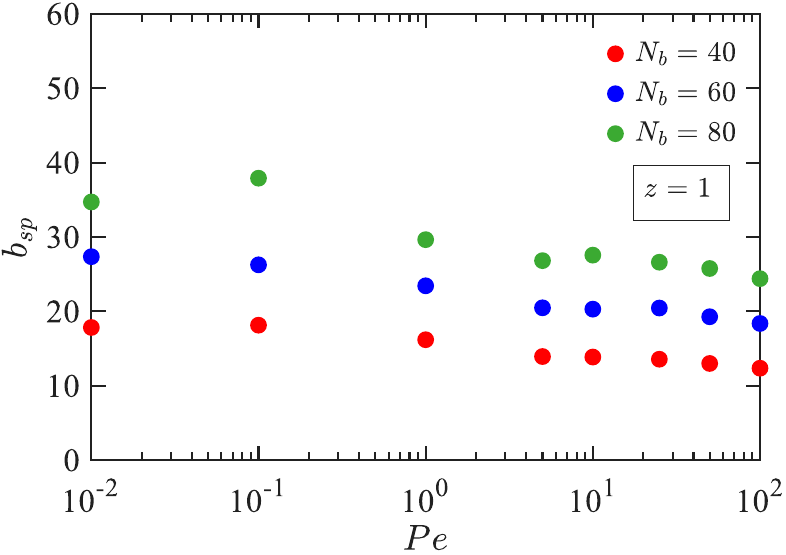} 
    & 
       \includegraphics[width=14cm,height=!]{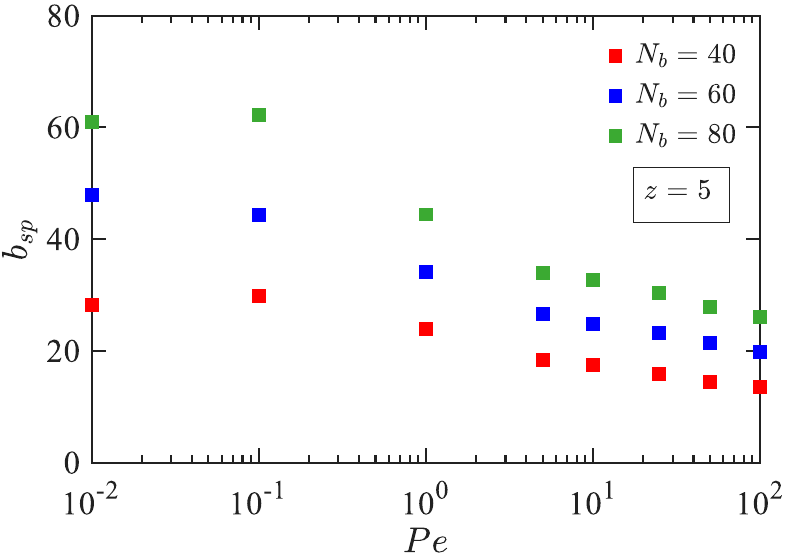} \\[5pt]
           \huge{(a)}  & 
       \huge{(b)} \\
       \includegraphics[width=14cm,height=!]{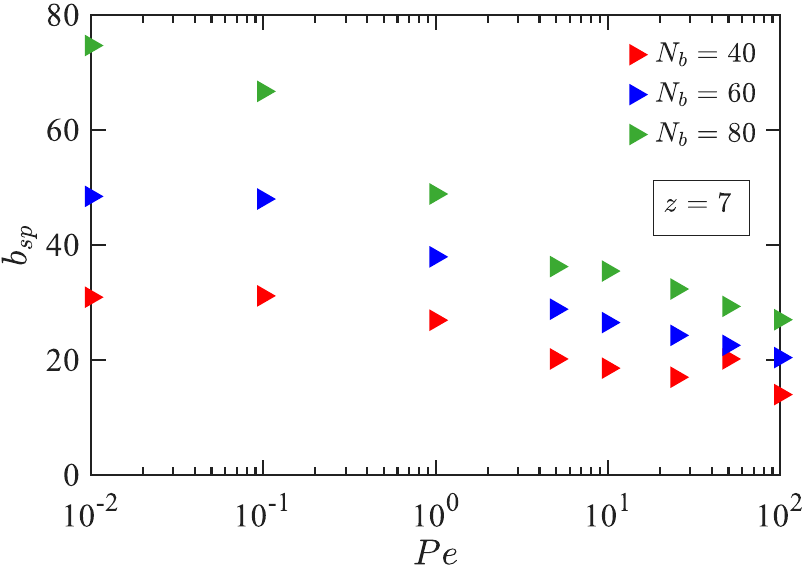} 
       &
       \includegraphics[width=14cm,height=!]{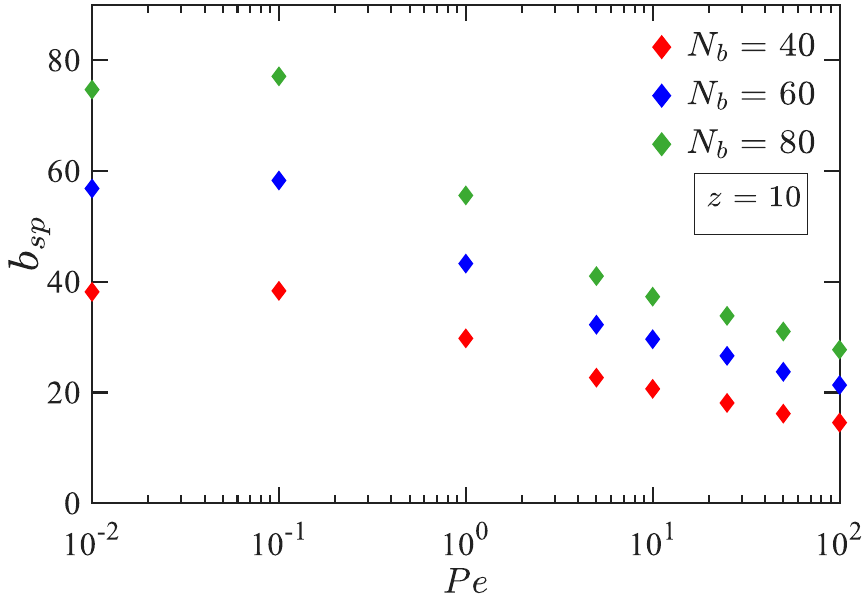} \\[5pt]
       \huge{(c)} & 
         \huge{(d)}  \\
  \end{tabular} }}
\caption{\small{Variation of asphericity parameter $b_{sp}$ as a function of P{\'e}clet number $Pe$ for different values of $N_b$, at $z=1,\,5,\,7$ and $10$. }} 
\label{fig:asphericity}
\end{figure*}

In Fig.~\ref{fig:shape_aniso}~(a), the shape anisotropy $\kappa^2$ is plotted as a function of P{\'e}clet number for different values of $N_b$, at solvent quality parameter $z=1,\,5,\,7$ and $10$. The variation of $\kappa^2$ with $Pe$ indicates that the shape anisotropy reaches an asymptotic value in the limit of large P{\'e}clet number (indicated by the dashed-solid horizontal lines in Fig.~\ref{fig:shape_aniso} (a)) for all the values of solvent quality parameter. Notably, as presented in Fig.~\ref{fig:shape_aniso} (b), the asymptotic value of shape anisotropy, $\kappa^2_{\infty}$, decreases with increasing $z$, following a power law scaling of the form, $\kappa_{\infty}^2=0.408\,z^{-0.039}$. The results suggest that the conformation of active polymers becomes increasingly symmetric with increasing P{\'e}clet number and is also affected by the solvent quality parameter $z$. Furthermore, the values of the asphericity parameter $b_{sp}$ is found to decrease with increasing $Pe$, at different values of $z$, as presented in Fig.~\ref{fig:asphericity} (a)-(d). The drop in the values of parameter $b_{sp}$ is found to be more sharper for the larger values of $z$ as compared to smaller $z$. Such a variation of the asphericity again indicates that the chain conformation becomes more symmetric and compact with increasing $Pe$ at all the values of solvent quality parameter in the thermal crossover regime.

%%%%%%%%%%%%%%%%%%%%%%%%%%%%%%%%%%%%%%%%%%%%%%%%%%%%%%%%%%%%%%%%%%%%%%%%%
\section{\label{sec:summary}Summary and Conclusion}
%%%%%%%%%%%%%%%%%%%%%%%%%%%%%%%%%%%%%%
In this study we have simulated tangentially active polymers in dilute solutions using bead-spring-chain model to investigate the effects of activity ($Pe$) on the polymer conformation in the thermal crossover regime between $\theta$ and athermal solvent conditions. Our analysis suggests that tangential activity leads to deswelling of polymer chains which reduces the effective solvent quality of active polymer solutions from an athermal condition to poor solvent, indicated by the decreasing value of exponent $\nu$. In particular, we show that there exists a unique value of P{\'e}clet number, $\bar{Pe}_{\theta}$, independent of the solvent quality parameter $z$, at which the polymer conformation represents a $\theta$ solvent condition. Subsequently, a swelling ratio for the radius of gyration of a chain relative to its value at the $\theta$-state is formulated for the active polymers and its variation is investigated as a function of a renormalized solvent quality parameter. Notably, the swelling behaviour is found to be identical to that of a passive polymer. Analysis of the shape anisotropy and asphericity parameter suggests an increasing conformational symmetry with P{\'e}clet number in the thermal crossover regime. Furthermore, the chain conformation statistics based on polymer blob theory presents a novel scaling relation for the thermal blob size of active polymers in terms of the solvent quality parameter and strength of activity. These results have important implications towards determining the equilibrium configurations of active polymer chains under different solution temperature and P{\'e}clet number, which would consequently help in predicting various dynamic and viscoelastic properties of active polymer solutions, such as diffusivity, viscosity, etc. Moreover, the present study provides a framework for analyzing and understanding the dynamics of active polymers which would be useful in systematic designing of active polymer based drugs and its targeted delivery, investigating motion of active filamentous micro-motors for various biophysical and biomedical applications.

\vspace{-0.2in}\section*{Acknowledgments}\vspace{-0.1in}

The work is partially supported by the startup research grant from IIT (ISM) Dhanbad with project no. FRS(204)/2023-2024/CHEMICAL. The authors acknowledge National Supercomputing Mission (NSM) for providing computing resources for the HPC System Param Himalaya, which is implemented by C-DAC and supported by the Ministry of Electronics and Information Technology (MeitY) and Department of Science and Technology (DST), Government of India. We appreciate the helpful conversations with R. Kailasham regarding implementation of the BD algorithm for active polymer chains.
 
%\mciteErrorOnUnknownfalse
\bibliography{APbibfile}
\bibliographystyle{unsrtnat}

\end{document}